\documentclass[12pt]{article}
\usepackage[utf8]{inputenc}

\usepackage{authblk}
\usepackage[left=1in,right=1in,top=1in,bottom=1in]{geometry}
\usepackage{placeins}
\usepackage{graphicx}
\usepackage{amsmath}
\usepackage[version=4]{mhchem}
\usepackage{siunitx}
\usepackage{longtable,tabularx}
\usepackage{array}
\usepackage{multirow}
\usepackage{color}
\usepackage{rotating}
\usepackage{verbatim}
\usepackage[normalem]{ulem}
\usepackage{epstopdf}

\usepackage[colorlinks=true,allcolors=blue]{hyperref}

\newcommand{\etal}{\emph{et al.}}

\begin{document}
\title{Destruction of wall-bounded vortices using synthetic jet actuators}

\author[1]{Frank A. Tricouros, \href{mailto:franktri@udel.edu}{franktri@udel.edu}.}
\affil[1]{Department of Mechanical Engineering, University of Delaware, Newark, Delaware, 19716}

\author[2]{Cameron Hoober}
\affil[2]{Mechanical and Aerospace Engineering, Illinois Institute of Technology, Chicago, IL 60616}

\author[3]{John C. Vaccaro}
\affil[3]{Department of Engineering, Hofstra University, Hempstead, New York, 11549}

\author[1]{Tyler Van Buren}

\setlength{\affilsep}{0.5em}
\renewcommand\Authfont{\normalsize}
\renewcommand\Affilfont{\small}
\date{}

\maketitle

\begin{abstract}
    We experimentally explore the effectiveness of a rectangular orifice synthetic jet actuator for wall-bounded vortex destruction. Vortex flows near a boundary often present unforeseen or undesired forcing on a neighboring surface due to the low pressure concentration within the vortex. Synthetic jets---primarily used for separation control, enhanced mixing, and induced turbulence---offer a unique strategy for vortex mitigation due to the unsteady flow at the region of the orifice disrupting the coherence of the oncoming flow. In a flat plate boundary layer, we test multiple jet orifice configurations, vortex lateral position relative to the orifice, and vortex sizes. We find that each jet was capable of reducing the incoming vortex rotational coherence up to 70\%. This disruption led to pressure recovery within the vortex wake region. The velocity wake of the vortex was more persistent (most jets produced a wake of their own) though some cases were capable of accelerating the fluid while maintaining moderate rotation reduction and pressure recovery. These results indicate that synthetic jets have the potential to mitigate a near wall vortex structure, particularly in scenarios where the position and size of the vortex are known.
\end{abstract}

\section{Introduction}\label{SEC:Intro}

Vortices are fundamental in fluid motion, playing a central role in aerodynamic performance \cite{LongitudinalControl}. Depending on the application, the influence of vortices can either be beneficial or detrimental \cite{mitchell2001research}. They can increase maneuverability \cite{hitzel2018enhanced}, energize the boundary layer \cite{VGLin}, and enhance heat transfer with improved mixing \cite{VGHeat}. However, vortices can also increase drag \cite{polhamus1966concept}, interfere with control authority \cite{soohoo1977aerodynamic}, and introduce vibrations and instabilities \cite{boersma2021experimental}. In scenarios where you need optical clarity through the fluid---e.g., downstream of a dome on an aircraft \cite{acarlar1987study}---vortex structures can introduce density gradients that impact imaging or laser-based technologies. The sensitivity of the flow globally to the presence of a vortex makes it ideal for flow control \cite{VorticalManagement}.
The suppression and destruction of coherent vortices remains a unique challenge. In this study, we investigate the use of synthetic jet actuators to disrupt and destabilize coherent wall-bounded streamwise vortices on a flat plate. 

Wall-bounded vortices are distinct both for their formation mechanism and their persistence. These elongated, coherent structures align their vorticity vector primarily with the mean flow, enabling momentum transfer across long distances \cite{crow1970stability}. Streamwise vortices can develop naturally from instability-driven processes, like how turbulent fluctuations can excite instability modes into coherent vortices \cite{robinson1991coherent}. For example, G\"ortler vortices arise when concave curvatures induce pressure gradients across the boundary layer, causing the flow to roll into streamwise vortex structures \cite{swearingen1987growth}. Similarly, when secondary flows, such as jets, are introduced into a crossflow, they can generate vortices\cite{khan2000vortex, VanBurenOrientation}. A streamwise vortex is often created, either inadvertently or deliberately, due to objects obstructing the local flow. Most commonly we see this with vortex generators---one of the most traditional flow control methods---that create regions of high and low pressure locally to induce circulation \cite{bartlett1973wind}. Other obstacles or protrusions, such as domes, can also shed vortices into the freestream flow \cite{acarlar1987study}. Regardless of how they are formed, these vortices often affect downstream flow mechanics. 

The negative effects of these streamwise vortices on vehicle performance are well documented. A common example is the induced drag generated by a wingtip vortex causing a long-persisting coherent wake downstream that can disrupt other vehicles or lifting surfaces \cite{spillman1978use, ActiveAttenuation}.  In the wake of bluff bodies, like missiles and submarines, asymmetric counter-rotating vortices lead to strong fluctuating moments, causing instabilities \cite{ashok_pitch}. On delta wings, vortex separation results in sudden losses of lift and pitching up motions \cite{RaoSegmented}. Vortices generated upstream of control surfaces severely impact their capabilities \cite{beresh2009interaction}. These examples emphasize the importance of exploring effective strategies to suppress or destroy the vortex’s structure, preserving or improving vehicle performance and stability.

Synthetic jets have long been used as a flow control device in various applications \cite{GlezerAmitay}, including separation prevention and mixing enhancement. We will demonstrate a more unique way to use synthetic jets for vortex destruction. Synthetic jets, also known as zero-net mass flux jets, generate their flow from oscillating diaphragms or membranes that sequentially inject and expel fluid through an orifice \cite{SmithGlezer}. Despite having zero-net mass flux, the alternating cycle produces a time-averaged jet with comparable momentum flux to that of a steady jet \cite{smith1998modification}. At the orifice exit, the unsteady jet produces a train of vortex rings that advect downstream \cite{GlezerVortex} and, depending on orifice geometry, can lead to downstream coherent vortex structure \cite{VanBurenJFM, wang2020interactions, TricourosJFM, TricourosEXIF}. In this work we adopt a rectangular exit geometry which can have controllable vortex production based upon orifice orientation and a larger width of impact of the flow when compared to a circular orifice of equal opening area. We selected synthetic jets as the control device in this application for two primary reasons: (1) with the right orifice geometry and orientation, they can significantly disrupt flows without creating any downstream vortex structure \cite{VanBurenOrientation}; and (2) we hypothesize that the inhaling and exhaling of the external flow into the jet will disrupt pre-existing flow vortex coherence. 
 
Despite extensive literature on synthetic jets for separation control and mixing, little is known about how they can be used for targeted vortex attenuation. Prior works highlight how synthetic jets can create vortices \cite{VanBurenJFM}, alter the trajectories of longitudinal vortices \cite{remneff2024control}, or combine and strengthen existing vortices \cite{van2015interaction}. Yet inversely, the problem of intentionally destabilizing or destroying pre-existing streamwise coherent vortices with synthetic jets remains a topic of exploration. A motivator for this work is a study that found unexpected interaction between a synthetic jet and a vortex generator \cite{van2015interaction}; where an attempt to enhance a vortex (also known as ``winding'') led to it being inadvertently destroyed and replaced by the flow field of the synthetic jet entirely.  Though the result was unintended, it revealed the potential of the actuator to be used more uniquely. By varying the pitch and skew angles of the jet orifice, we demonstrate how the synthetic jet interacts with the vortex core to destabilize and accelerate the vortex breakdown. The results reveal a measurable attenuation of vortex strength with varying levels of pressure recovery.

\section{Experimental methods}\label{SEC:Methods}

\begin{figure}
    \centering
    \includegraphics[width=\linewidth]{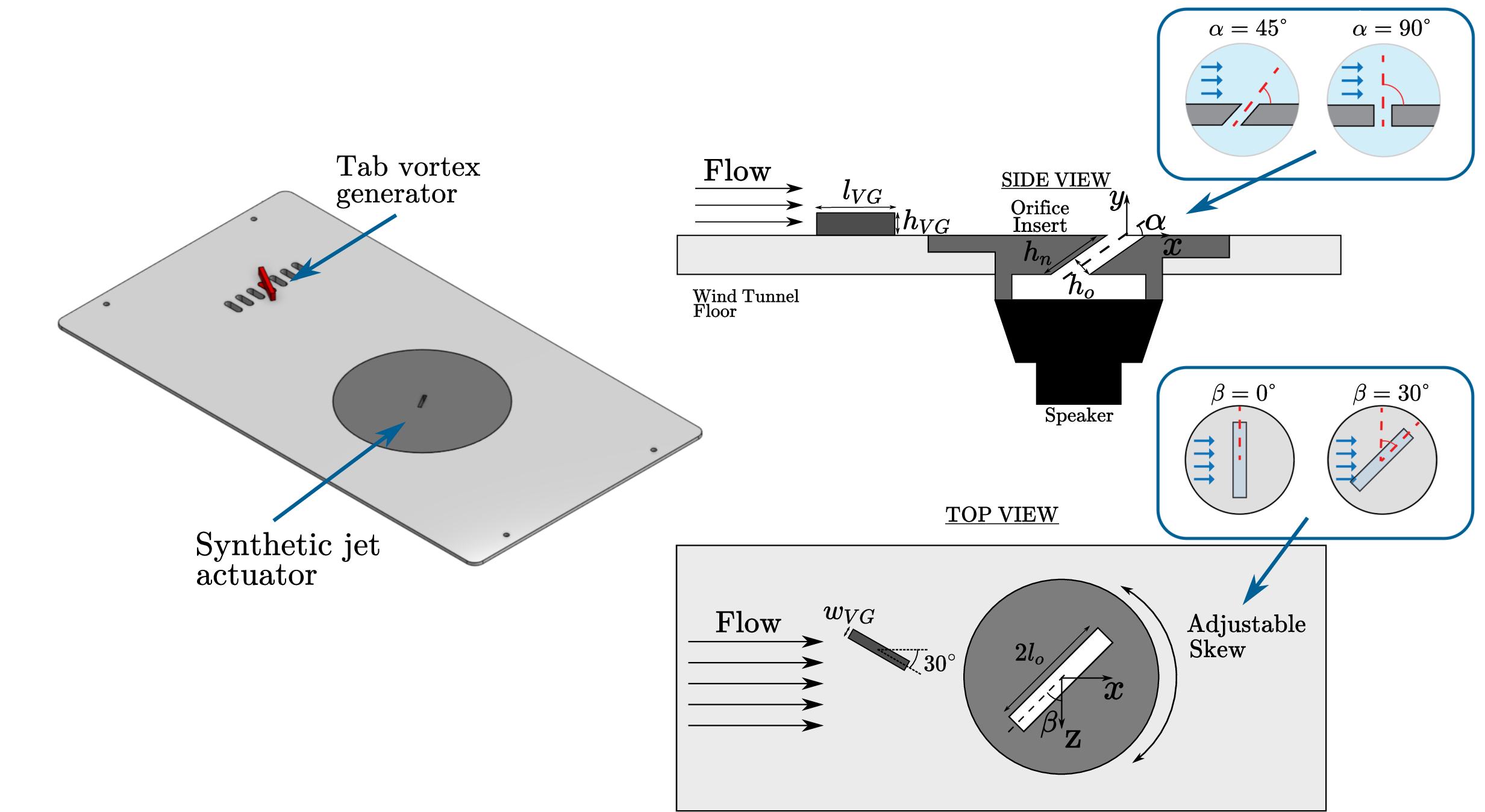}
    \caption{Schematic of the side and top view of the wind tunnel insert with variable vortex generator and synthetic jet positioning.}
    \label{fig:expSetup}
\end{figure}

The dataset was collected in an Aerolab EWT open return suction wind tunnel. The wind tunnel featured a turbulence intensity of less than 0.25\% and had a cross-sectional area of 0.305 $\times$ 0.305 m$^2$ with a test section length of 0.61 m. Throughout the experiments, a freestream velocity of $U_\infty = 10$ m/s was maintained, monitored using a pitot-static tube connected to a Scanivalve DSA3217 multimanometer. The uncertainty of the multimanometer was $2.5\times10^{-3}$ kPa, corresponding to velocity errors of $\pm 0.2$ m/s. Acrylic side walls were installed to provide optical access for our measurement techniques. The floor of the tunnel was replaced with our experimental apparatus.

We elected to use rectangular tab-style vortex generators to create coherent and well defined wall-bounded streamwise vortex structures. Two vortex generators were used, varying only the height between them, $h_{VG} = $ 7.5 or 10 mm. All vortex generators featured a length $\ell_{VG} = 40$ mm, width $w_{VG} = 4$ mm, and were angled $\theta = 30^\circ$ relative to the oncoming flow. The vortex generator centroids were located 210 mm upstream of the synthetic jet orifice slot centroid. Our wind tunnel floor insert had variable spanwise (z-direction) cutouts where a spanwise offset of 40 mm lead to the vortices passing directly over the center of the synthetic jet apparatus. When not in use, these cutouts were capped with blanks and taped to ensure no flow leakage. We have also explored other positions of the vortex generator to gauge the effects of vortex proximity to the actuator.

Visaton SC 8N speakers were used to generate our synthetic jets. They were attached to an orifice insert plate that sat on the wind tunnel floor. The speakers were driven with a 220 Hz sinusoidal voltage signal generated using a Siglent SDG 1032X arbitrary waveform generator. The waveform output was amplified using a Fosi Audio TB10D speaker amplifier. The root mean square voltage output was monitored using a Keysight U1242B Multimeter (accuracy of $0.09\%$) to ensure repeatable settings. We tested the jets with three separate velocity settings, corresponding to $V_{RMS} = 5$, 10, or 15.3 V. However, we prioritized maximizing jet interaction and collected most of the data at $V_{RMS} = 15.3$V. The higher voltage setting resulted in blowing ratios of $C_b = 1.0$ and $C_b = 1.3$ for the pitched $\alpha=45^\circ$ and wall-normal $\alpha = 90^\circ$ jets, respectively. The orifice plates had a width of $h_o$ = 2 mm, slot length $2l_o = 18$ mm, and neck height $h_n = 7/\sin{\alpha}$ mm. Three orifice pitch angles were explored, $\alpha = 45^\circ, 90^\circ, $ and $135^\circ$ by swapping out the orifice inserts. The orifice skew of $\beta = 0^\circ$ and $30^\circ$ was adjusted by rotating and locking in the orifice insert.

A commercial LaVision Stereoscopic Particle Image Velocimetry (SPIV) system was used to collect two-dimensional, three-component velocity field planes at multiple streamwise locations. Imager SX 4M cameras were used to capture data, they had 12-bit CCDs with a resolution of 2330 $\times$ 1750 pixels. The cameras were mounted onto Scheimpflug lens adapters and equipped with Nikon 105 mm focal length lenses with 532 nm $\pm$ 10 nm band-pass filters. The cameras were positioned on opposite sides of the wind tunnel and angled $40^\circ$ relative to the streamwise direction, $80^\circ$ relative to each other.  A Rocket Smoke Machine was used to seed the flow, producing particles ranging from 0.2 to $0.3 \mu$m in diameter. A Quantel Evergreen 145 mJ/pulse Nd:YAG dual cavity pulsed laser was used to illuminate the seed particles. LaVision DAVIS software was used to process the data. A cross-correlation of successive pairs of images technique was used to calculate the velocity vectors. Time-averaged data consisted of 500 image pairs taken with time steps of $25$ $\mu$s. The frequency of data acquisition for successive image pairs was 5.583 Hz to avoid phase-locking the data acquisition to the unsteady jet actuation frequency. For phase-locked data, 500 image pairs were taken with time steps of $5 \mu$s. Phase-locked data increments were every 45$^\circ$ from $0-315^\circ$. A multi-pass method was used to process the images where the first pass used $48 \times 48$ pixel interrogation windows with $50 \%$ overlap, followed by two passes with $32 \times 32$ pixel windows with $50 \%$ overlap. To nondimensionalize length scales and velocity scales we use the orifice width and the freestream velocity, respectively. 

\section{Results and discussion}\label{SEC:Results}
Using synthetic jets, we aim to disrupt a streamwise wall-bounded vortex in flat plate boundary layer flow. One primary cause of disturbance of a streamwise vortex in various applications is the low pressure induced by the vortex rotation, leading to unplanned body forces on the vehicle or lifting surface local to the vortex. The source of the low pressure is rooted in the azimuthal velocity of the vortex, which we can see through the fluid conservation of momentum equations in differential and cylindrical form
\begin{equation}
    \frac{1}{\rho}\frac{\partial P}{\partial r} = \frac{u_\theta^2}{r}.
\end{equation}
Here the flow is assumed to be two-dimensional with no radial velocity, body force, or time dependence. An azimuthal velocity $u_\theta$ leads to a radial gradient of pressure $\frac{\partial P}{\partial r}$, ultimately causing the core of a vortex to have concentrations of low pressure. It follows that, if we can disrupt the azimuthal velocity of the vortex, we can reduce the vortex's influence.

\begin{figure}
    \centering
    \includegraphics[width=\linewidth]{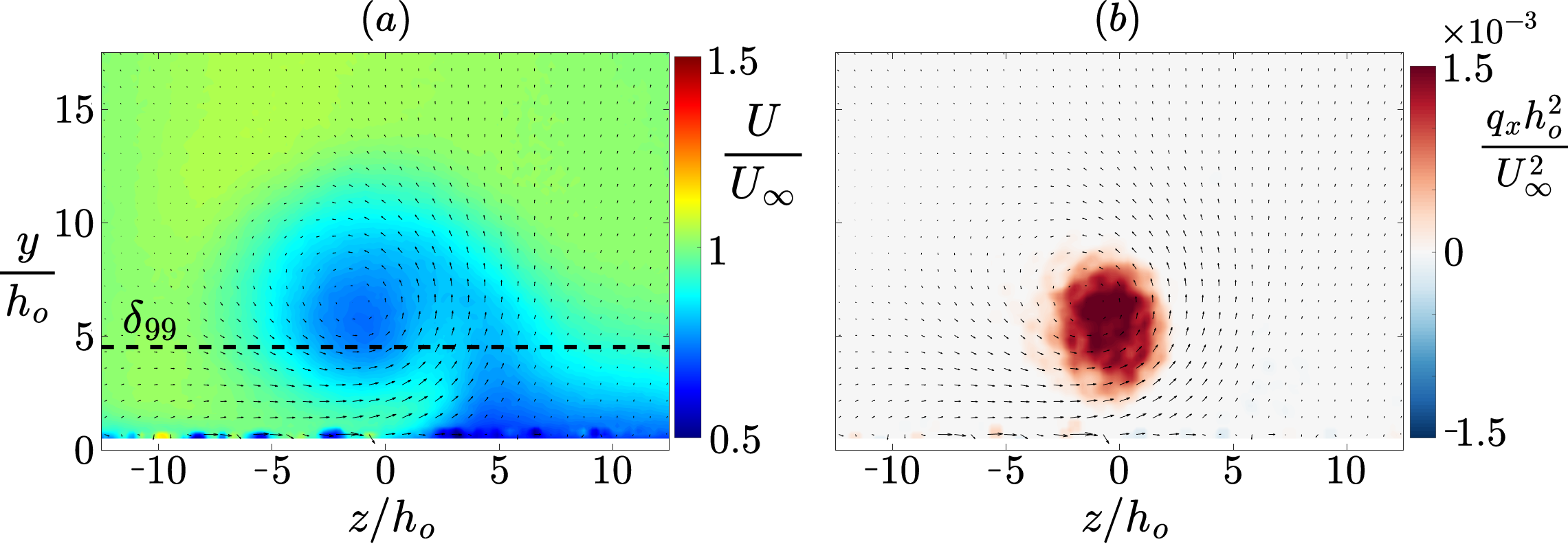}
    \caption{Characterization of the vortex from the vortex generator including the velocity (a) and q-criterion (b) fields. Measurements made at $x/h_o=5$ downstream of a $l_\text{VG}/h_o = 20$ \& $h_\text{VG}/h_o = 3.75$ vortex generator. The boundary layer height is referenced.}
    \label{fig:vortexCharacterization}
\end{figure}

\subsection{Characterization of flow control devices}
\noindent To begin our analysis, we characterize the generated vortex without a synthetic jet present. We use a modified version of the $\mathcal{Q}$-criterion field to assess the rotational structure of the vortex isolated from other sources of vorticity, quantified by
\begin{equation}
    q_x = -\frac{1}{2} \frac{\partial v}{\partial y}\frac{\partial v}{\partial y} - \frac{1}{2}\frac{\partial w}{\partial z}\frac{\partial w}{\partial z} - \frac{\partial v}{\partial z}\frac{\partial w}{\partial y}.
    \label{eq:q_criterion}
\end{equation}
Though traditionally $\mathcal{Q}$-criterion is a total quantity, here we calculate it only from the in-plane components of velocity, $v$ and $w$, thus denoting the criterion with a $x$ subscript and terming it $q_x$ or q-criterion. It is also given the sign of the local streamwise vorticity, so we can identify vortex structures of opposing rotation directions. Figure \ref{fig:vortexCharacterization} shows the velocity and q-criterion fields for the vortex generated by a tab-style vortex generator measured $x/h_o=5$ downstream. In the velocity field there is a clear rotational structure with a concentration of streamwise velocity deficit inside the structure. The baseline (vortex free) boundary layer is denoted in height by $\delta_\text{99}$ (marking where the baseline spanwise-averaged velocity is 99\% of the freestream). On the left side of the vortex (where the flow is toward the wall), high momentum flow is pushed into the boundary layer. This is exchanged with the low momentum fluid, which is subsequently exhausted to the freestream on the right side (where flow is away from the wall). The vortex is roughly 2-3 times the height of the local flat-plate boundary layer. The q-criterion shows a concentration of rotational structure in the center of the vortex, successfully identifying the vortex core region. Throughout the work we will use this as our baseline vortex, unless otherwise stated. 

\begin{figure}
    \centering
    \includegraphics[width=\linewidth]{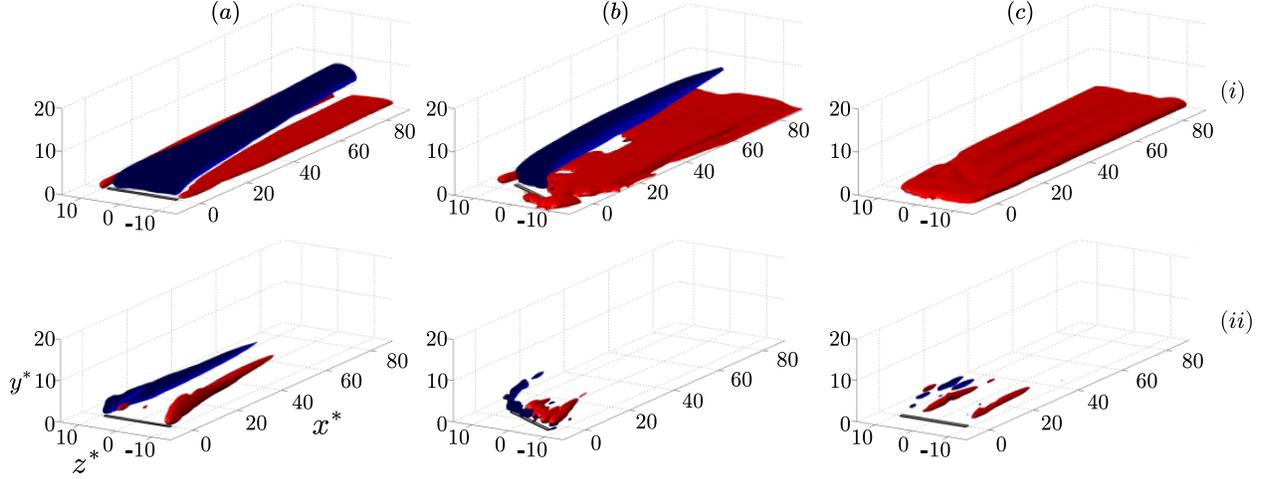}
    \caption{Characterization of three of the synthetic jet orientations: (a) $\alpha = 90^\circ$ \& $\beta = 0^\circ$, (b) $\alpha = 90^\circ$ \& $\beta = 30^\circ$, and (c) $\alpha = 45^\circ$ \& $\beta = 0^\circ$ at a blowing ratio of $C_b = 1.0$. Isosurfaces of the change in total velocity (i) and vortex structure via streamwise q-criterion (ii) are shown for each. Figures replicated from \cite{VanBurenOrientation}.}
    \label{fig:sjCharacterization}
\end{figure}

Next, we can characterize the synthetic jet flow field in absence of a pre-existing wall-bounded vortex. It is important to consider the vortex structure generated by the synthetic jet itself, as we would like to avoid a scenario where we destroy one vortex to make multiple others. As was shown in Van Buren \etal \cite{VanBurenOrientation} changing the orifice orientation of a rectangular orifice synthetic jet dramatically impacts the resulting velocity and vorticity field. Figure \ref{fig:sjCharacterization} shows the change in velocity relative to the baseline flat plate boundary layer and q-criterion fields for a synthetic jet with a rectangular orifice \cite{VanBurenOrientation}. We target three specific orientations: $\alpha = 90^\circ$ \& $\beta = 0^\circ$, $\alpha = 90^\circ$ \& $\beta = 30^\circ$, and $\alpha = 45^\circ$ \& $\beta = 0^\circ$ as potential vortex destruction candidates (data originally from Van Buren \etal \cite{VanBurenOrientation}). The two latter cases produce little to no vortex structure downstream, while still energizing the boundary layer with relatively small wake regions in the freestream. This is because when the jet pitch angle $\alpha$ is tilted forward, less streamwise vorticity is produced at the orifice; and when the jet skew angle $\beta$ is such that the orifice is not aligned with the freestream, there is also less produced streamwise vorticity. In both cases, this leads to less coherent streamwise vortex structure downstream. The more traditional wall-normal, perpendicular orifice $\alpha = 90^\circ$ \& $\beta = 0^\circ$ case is included which produces a clear streamwise vortex pair, and larger velocity deficit region in the wake. This is to test whether using an actuator which generates its own vortex structure impacts the destruction of the pre-existing vortex.

Along with the three synthetic jet cases shown above, we also include a novel orientation that has not yet been explored in past literature, $\alpha = 135^\circ$ \& $\beta = 0^\circ$, which should dramatically disrupt the velocity field by vectoring the jet against the flow. 

\begin{figure}
    \centering
    \includegraphics[width=\linewidth]{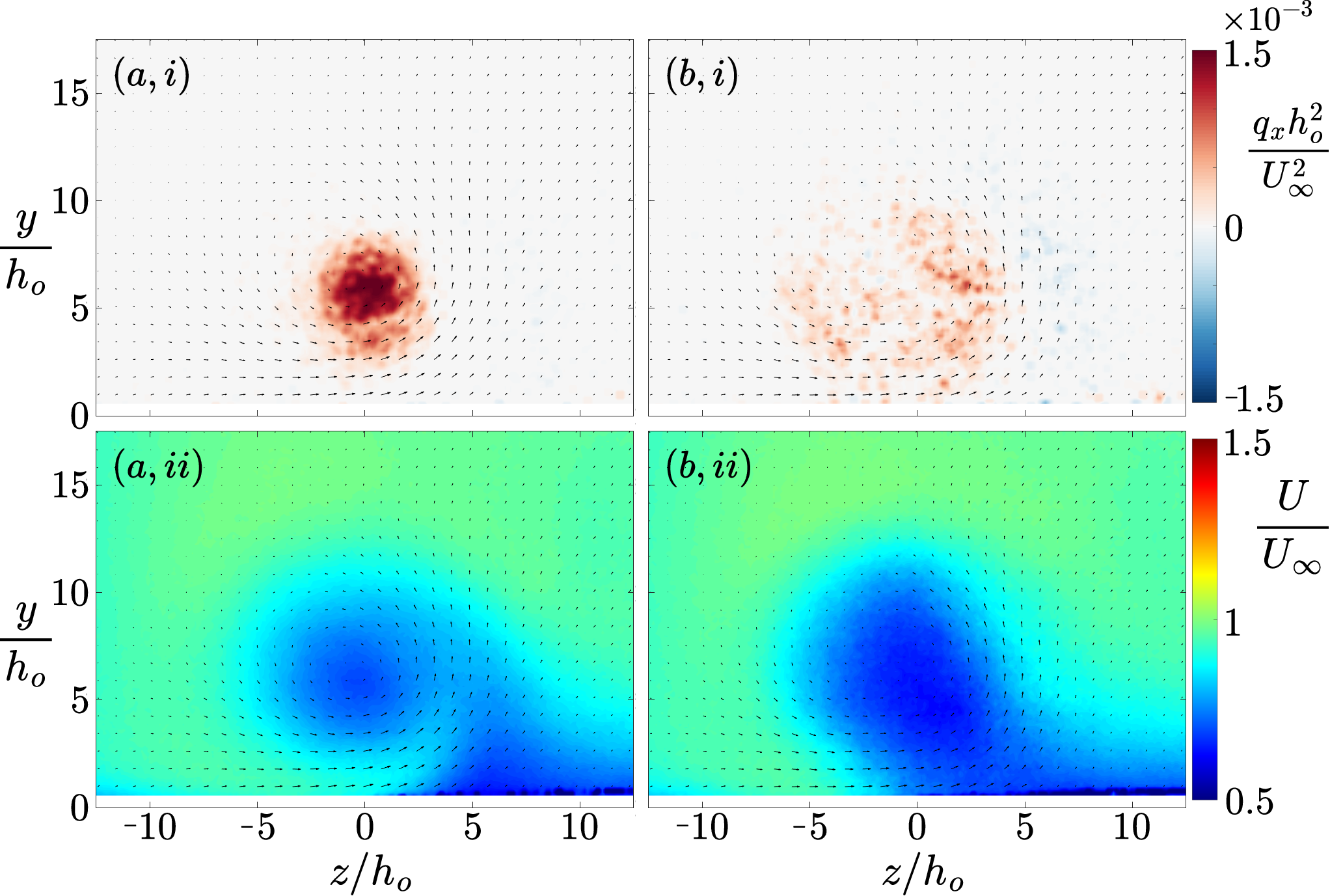}
    \caption{Impact of the synthetic jet on an upstream generated vortex, represented by q-criterion (i) and velocity (ii) fields, comparing the actuator being off (a) and turned on (b) measured at $x/h_o = 20$. The synthetic jet has orientation $\alpha = 90^\circ$ \& $\beta = 30^\circ$ and blowing ratio $C_b = 1.3$, and the vortex generator with $l_\text{VG}/h_o = 20$ \& $h_\text{VG}/h_o = 3.75$ in middle position.}
    \label{fig:singleCase}
\end{figure}

\subsection{Ideal case for vortex position}
\noindent With the vortex generator at the middle position on the wind tunnel insert described in section \ref{SEC:Methods}, we produced a vortex that passed directly over the orifice of the synthetic jet. This scenario represents an idealized condition for control authority. Figure \ref{fig:singleCase} shows the impact of a synthetic jet on a pre-existing vortex compared to the baseline case without jet actuation. The synthetic jet here is the wall-normal $\alpha = 90^\circ$ with skew angle \& $\beta = 30^\circ$ and a jet strength of $C_b = 1.3$. We see a clear impact of the jet on both the velocity and vortex structure. First, the coherence of the vortex---displayed in the q-criterion---is nearly completely diminished by the jet within $x/h_o = 20$ downstream. (The quivers on the q-criterion field representing the in-plane velocities do indicate some footprint of the azimuthal velocity structure). The streamwise velocity deficit originally present in the center of the vortex is still evident after actuation, despite the loss of vortex coherence. This is not unexpected, this synthetic jet did exhibit a wake in the original characterization so there is no logical reason the jet would accelerate a pre-existing wake. 

\begin{figure}
    \centering
    \includegraphics[width=\linewidth]{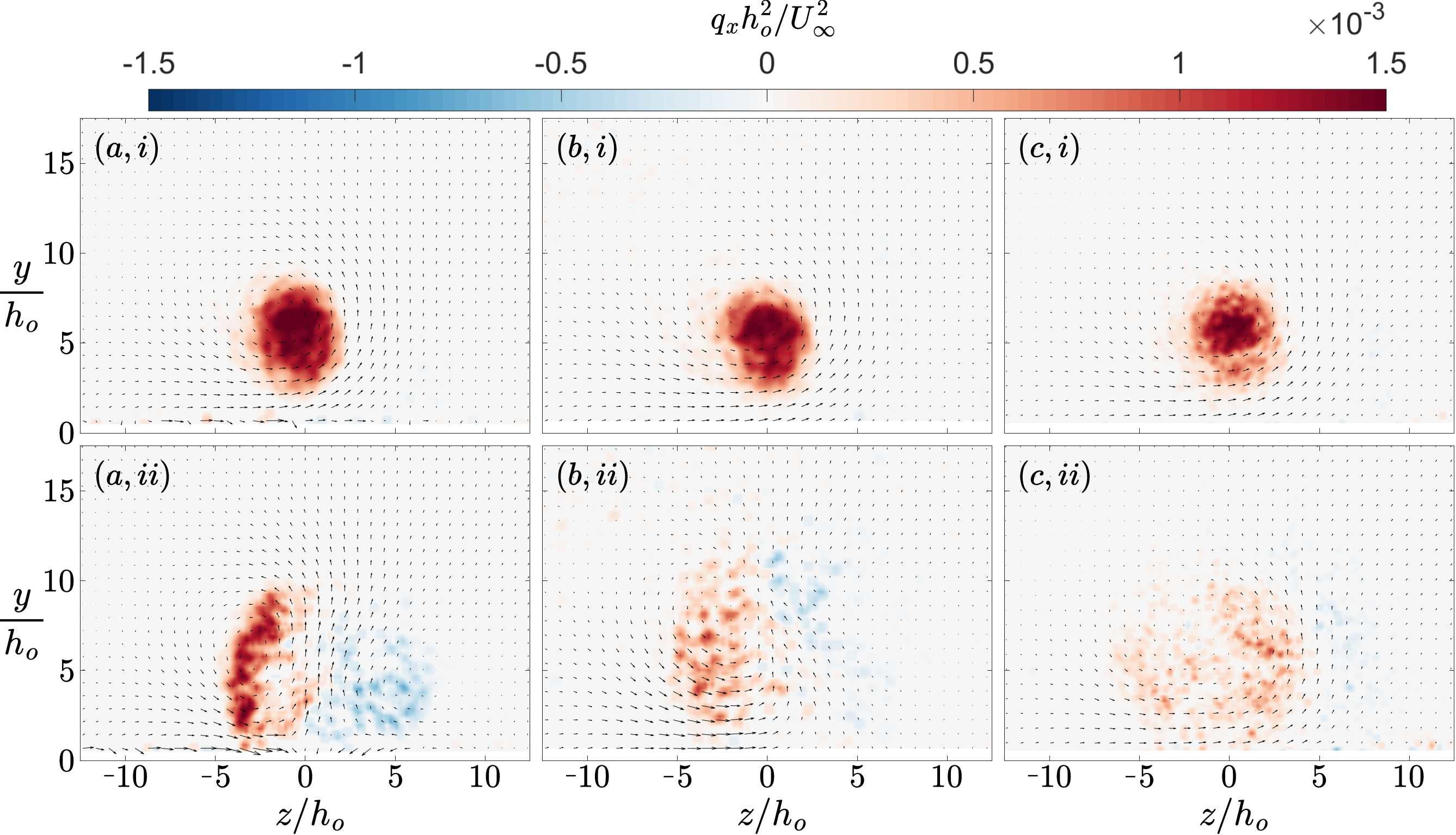}
    \caption{Downstream development of the vortex development with the actuator off (i) and on (ii) with measurements made at (a) $x/h_o= 5$, (b) $x/h_o= 10$, and (c) $x/h_o= 20$ downstream.  The synthetic jet has orientation $\alpha = 90^\circ$ \& $\beta = 30^\circ$ and blowing ratio $C_b = 1.3$, and the vortex generator with $l_\text{VG}/h_o = 20$ \& $h_\text{VG}/h_o = 3.75$ in the middle position.}
    \label{fig:downstreamDev}
\end{figure}

It is important to assess the downstream development of the effect to gauge how rapidly the vortex is mitigated. In figure \ref{fig:downstreamDev}, we show the vortex structure for the baseline vortex flow and actuated cases at multiple streamwise locations. In the un-actuated case, the vortex rotational structure gradually decays from $5 \geq x/h_o \geq 20$ naturally. However, due to the jet impact on the vortex, the vortex loses a majority of the coherence within $x/h_o=5$, leaving less concentrated regions of positive or negative $q_x$. This then rapidly diminishes, leaving behind almost no rotational evidence of a vortex by the end of the interrogation region. It is quite clear that this synthetic jet orientation and strength---admittedly in an ideal setting---can immediately destroy a local vortex that is comparable in size (i.e., the vortex diameter is on the order of the jet orifice length). 

\begin{figure}
    \centering
    \includegraphics[width=\linewidth]{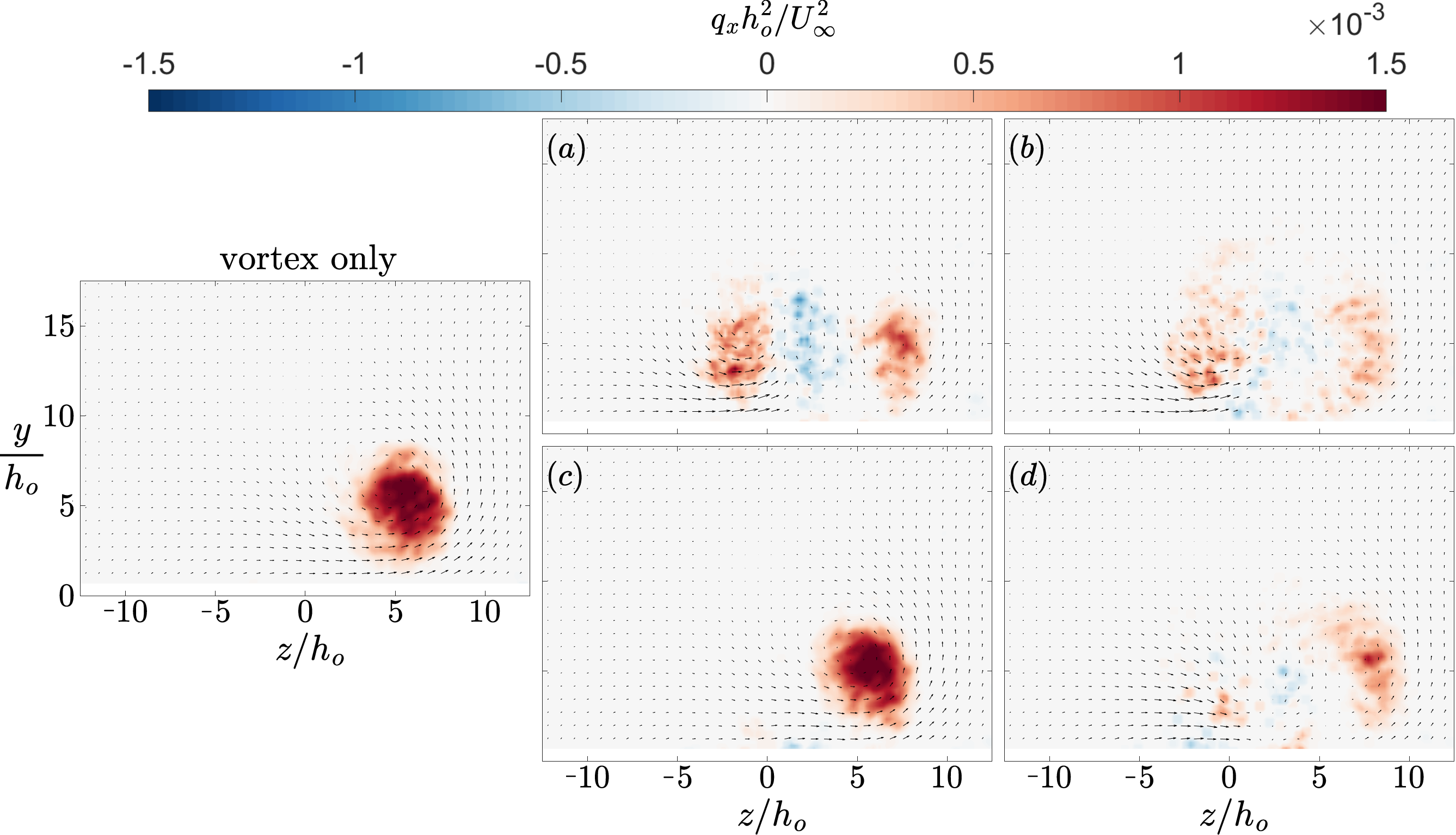}
    \caption{Impact of synthetic jet actuator on the q-criterion of an upstream generated vortex with (a) $\alpha = 90^\circ$ \& $\beta = 0^\circ$, (b) $\alpha = 90^\circ$ \& $\beta = 30^\circ$, (c) $\alpha = 45^\circ$ \& $\beta = 0^\circ$, and (d) $\alpha = 135^\circ$ \& $\beta = 0^\circ$ and blowing ratios $C_b = 1.3$ for the $\alpha = 90^\circ$ jets and $C_b = 1.0$ for pitched $\alpha = 45, 135^\circ$ jets, measured at $x/h_o = 10$. The baseline vortex flow field without jet actuation is shown on the left. The vortex generator had $l_\text{VG}/h_o = 20$ \& $h_\text{VG}/h_o = 3.75$ in the right position.}
    \label{fig:qMultSJ}
\end{figure}

Next, we determine the best jet orifice configuration for vortex destruction. We consider the four orifice orientations characterized in figure \ref{fig:sjCharacterization} plus the additional case that blows upstream against the flow, $\alpha = 135^\circ$ \& $\beta = 0^\circ$. Figure \ref{fig:qMultSJ} shows the impact the synthetic jets have on a pre-existing vortex via the q-criterion measured at $x/h_o=20$ downstream. While all cases have considerable impact on the vortex, there are varying degrees of effectiveness. The normal and perpendicular orifice case ($\alpha = 90^\circ$ \& $\beta = 0^\circ$) yields three distinct vortex regions. The first two are made up of the counter-rotating vortex pair associated with this orientation. The third shown is the diminished vortex due to the interaction between the jet clockwise vortex and the vortex generator counter-clockwise vortex structure. This is likely because in this configuration the vortex structure produced by the synthetic jet is strongest (see figure \ref{fig:sjCharacterization}). The pitched case with a perpendicular orifice ($\alpha = 45^\circ$ \& $\beta = 0^\circ$) has the weakest impact---this jet has the least virtual blockage (blockage due to the jet flow structure) of all the cases and is generally used for flow acceleration. Both the wall normal skewed case ($\alpha = 90^\circ$ \& $\beta = 30^\circ$) and the jet vectored into the flow ($\alpha = 135^\circ$ \& $\beta = 0^\circ$) have similarly strong impacts on the vortex structure, indicating that blockage of the vortex is important while also ensuring the synthetic jet is not producing its own vortex structure. It appears that vectoring the jet into the flow is not necessary, as less extreme cases have similar impact.

\begin{figure}
    \centering
    \includegraphics[width=\linewidth]{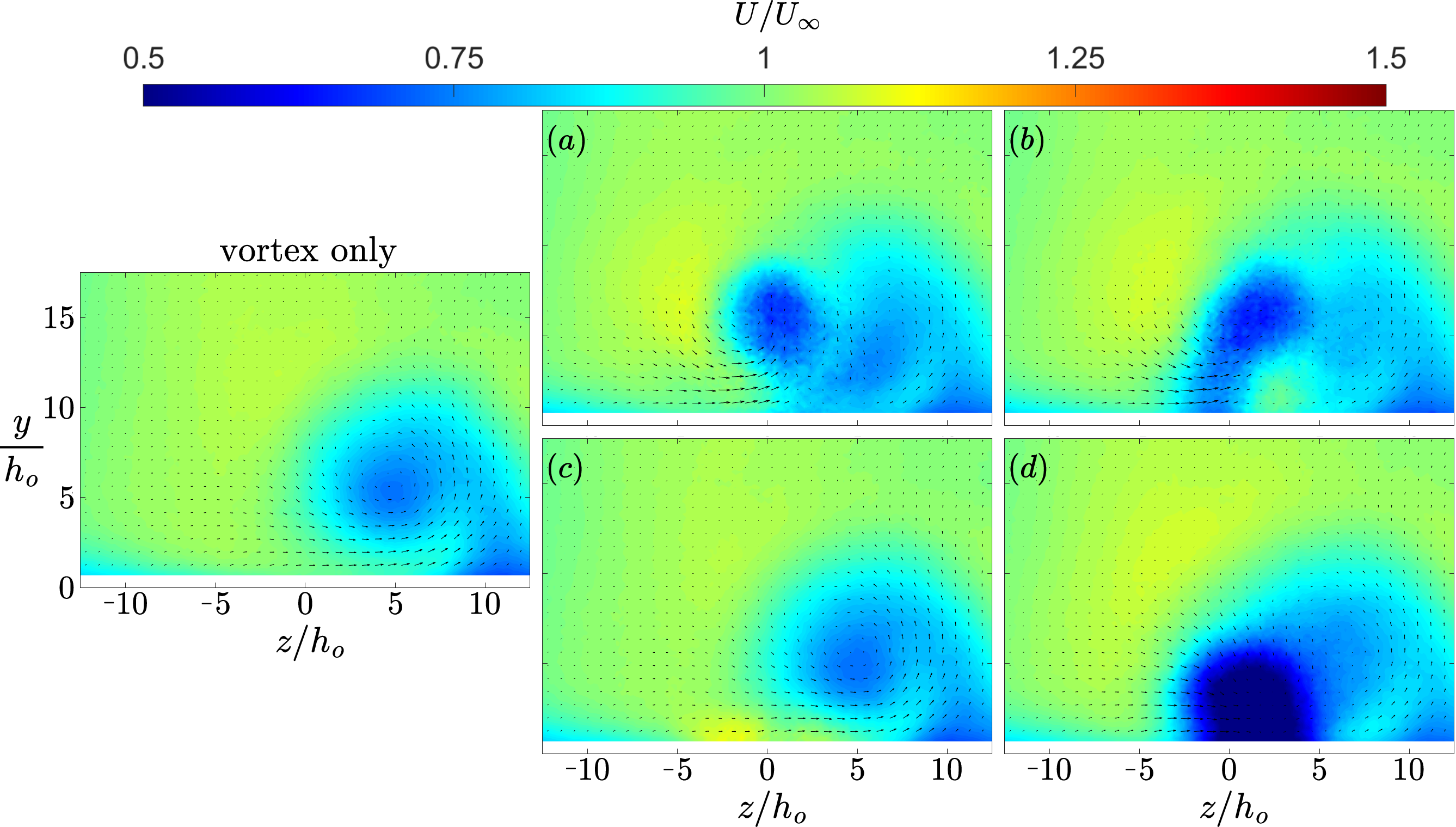}
    \caption{Impact of synthetic jet actuator on the velocity field of an upstream generated vortex with (a) $\alpha = 90^\circ$ \& $\beta = 0^\circ$, (b) $\alpha = 90^\circ$ \& $\beta = 30^\circ$, (c) $\alpha = 45^\circ$ \& $\beta = 0^\circ$, and (d) $\alpha = 135^\circ$ \& $\beta = 0^\circ$ and blowing ratios $C_b = 1.3$ for the $\alpha = 90^\circ$ jets and $C_b = 1.0$ for pitched $\alpha = 45, 135^\circ$ jets, measured at $x/h_o = 10$. The baseline vortex flow field without jet actuation is shown on the left. The vortex generator had $l_\text{VG}/h_o = 20$ \& $h_\text{VG}/h_o = 3.75$ in the right position.}
    \label{fig:veloMultSJ}
\end{figure}

Consider the velocity field for the same cases from figure \ref{fig:veloMultSJ}. In all cases, the wake of the vortex is similar to or worsened by the presence of the synthetic jet. The only case to feature velocity acceleration in the wake region is the case where the jet is vectored along the flow direction ($\alpha = 45^\circ$ \& $\beta = 0^\circ$), likely because this jet, on its own, directly accelerates the flow near the wall. However, there is no noticeable velocity deficit reduction for the wake region associated with the vortex itself. The opposite case, where the jet is vectored against the flow ($\alpha = 135^\circ$ \& $\beta = 0^\circ$), shows a much larger deficit region than the baseline vortex only case. This is to be expected because a jet injecting against the stream will created a considerable velocity deficit. Overall, in the cases we studied there were no combinations that were effective at mitigating both the rotational structure and the wake, though it is not immediately clear if there is negative consequence to a wake remaining after vortex destruction.

As we discussed earlier, the impact of the vortex on a neighboring surface is due to the low pressure within the vortex. Naturally, we evaluate the resulting pressure field consequences of our vortex destruction methods. The pressure field is calculated from the velocity field, with the pressure Poisson equation \ref{eq:PoissonPressure} for an incompressible fluid, following Barba \etal \cite{barbagroup_cfd_intro}
\begin{equation}
    \label{eq:PoissonPressure}
    \frac{\partial^2 P}{\partial x^2} + \frac{\partial^2 P}{\partial y^2} = -\rho \left(\frac{\partial u}{\partial x}\frac{\partial u}{\partial x} + 2\frac{\partial u}{\partial y}\frac{\partial v}{\partial x} + \frac{\partial v}{\partial y}\frac{\partial v}{\partial y}  \right).
\end{equation}
To solve this, we start with an initial pressure field of zeros and iteratively solve utilizing O($h^2$) finite difference schemes. The solution field was deemed converged when the change in pressure field between iterations varied by less than  $1 \times 10^{-6}$ Pa. Homogeneous Neumann boundary conditions were used at the edges of our domain.

\begin{figure}
    \centering
    \includegraphics[width=\linewidth]{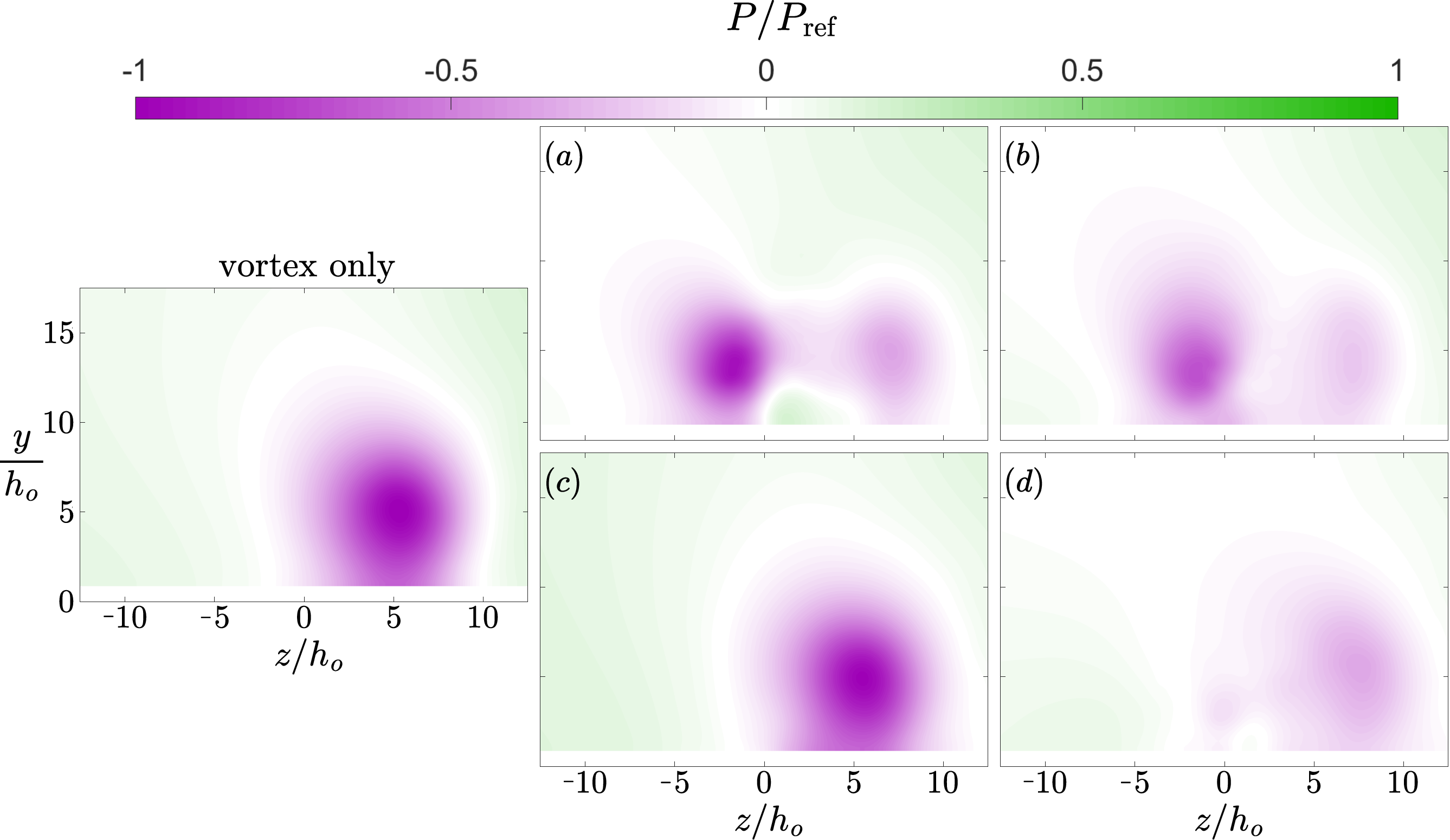}
    \caption{Pressure field comparison for a synthetic jet actuator interacting with an upstream generated vortex with (a) $\alpha = 90^\circ$ \& $\beta = 0^\circ$, (b) $\alpha = 90^\circ$ \& $\beta = 30^\circ$, (c) $\alpha = 45^\circ$ \& $\beta = 0^\circ$, and (d) $\alpha = 135^\circ$ \& $\beta = 0^\circ$ and blowing ratios $C_b = 1.3$ for the $\alpha = 90^\circ$ jets and $C_b = 1.0$ for pitched $\alpha = 45, 135^\circ$ jets, measured at $x/h_o = 10$. The baseline vortex flow field without jet actuation is shown on the left. The vortex generator had $l_\text{VG}/h_o = 20$ \& $h_\text{VG}/h_o = 3.75$ at $z/h_o = 6.75$.}
    \label{fig:PressureFieldDifference}
\end{figure}

Contours of the calculated pressure field are shown in figure \ref{fig:PressureFieldDifference}. To better visualize the relative impact each of the jet orientations has on the pressure field, we normalize by $P_\text{ref}$, a reference pressure which corresponds to the peak of the absolute value in the baseline vortex generator pressure field. Overall, most jets were able to weaken the negative pressure field associated with the vortex core, except for $\alpha = 45^\circ$, $\beta = 0^\circ$. However, the jets pitched normal to the floor, $\alpha = 90^\circ$, created secondary pressure structures at this vortex generator lateral position. In the case of the jet angled into the crossflow, $\alpha = 135^\circ$, $\beta = 0^\circ$, the pressure field was severely diminished in strength without generating secondary structures. It is a bit counter-intuitive that the orientation which produced the strongest velocity wake has the most pressure recovery. 

\subsection{Impact of vortex size and position}
\noindent Now that we understand the impact of a synthetic jet in an idealized case, we can start to consider the limitations of the jet effectiveness as the proximity and size of the pre-existing vortex change. Note that throughout this section we vary the vortex properties and control with our four representative orifice orientations for velocity and vortex field analysis.

\begin{figure}
    \centering
    \includegraphics[width=0.93\linewidth]{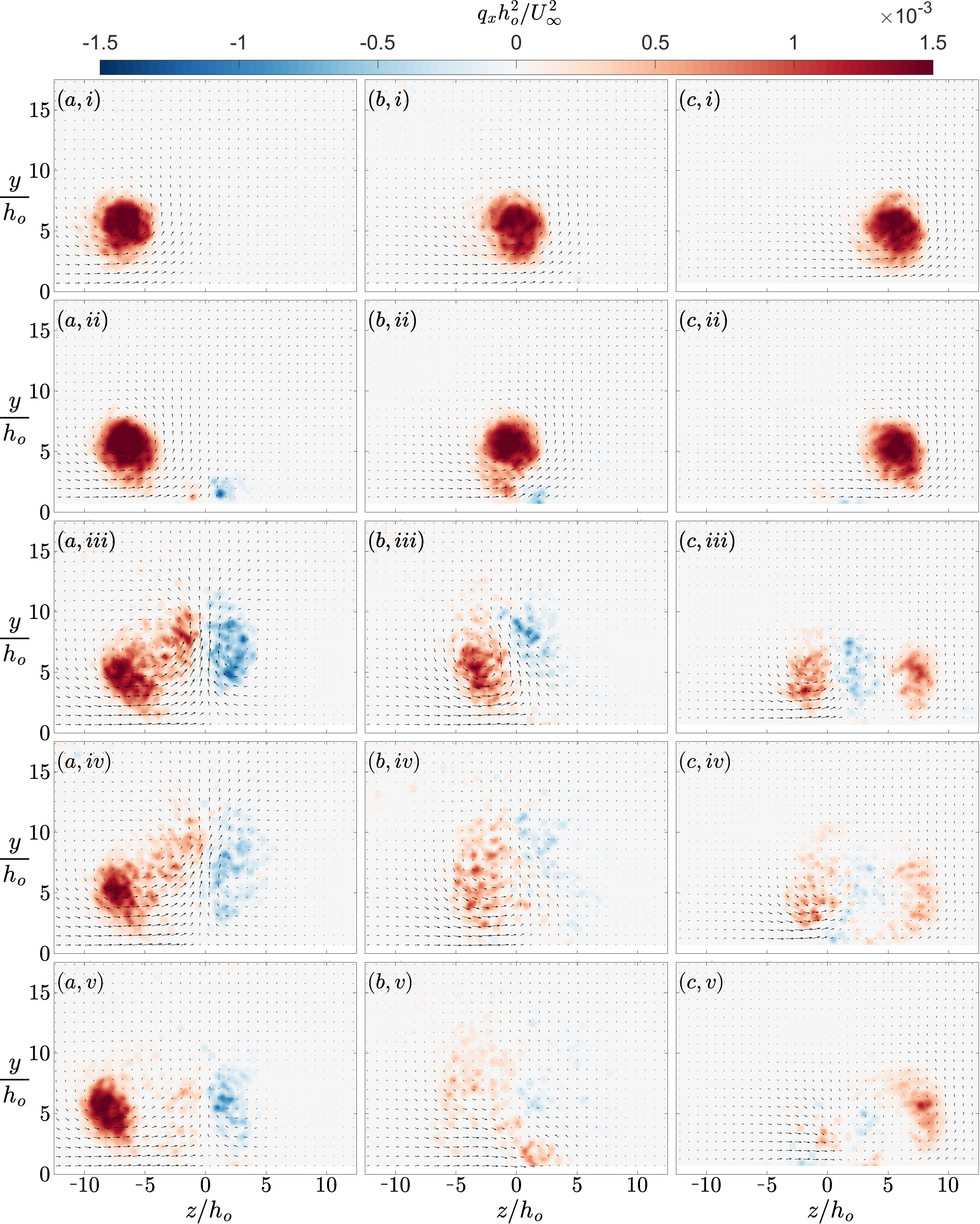}
    \caption{Vortex structure flow fields showing the impact of lateral vortex position on control effectiveness for a vortex positioned (a) $z/h_o = -6.75$, (b) $z/h_o = 0$, and (c) $z/h_o = 6.75$ relative to the synthetic jet centroid. Baseline flow fields of just the vortex generators are shown in row (i). Synthetic jet orientations were (ii) $\alpha = 45^\circ$ \& $\beta=0^\circ$, (iii) $\alpha = 90^\circ$ \& $\beta=0^\circ$, (iv) $\alpha = 90^\circ$ \& $\beta=30^\circ$, and (v) $\alpha = 135^\circ$ \& $\beta=0^\circ$. All flow fields shown are at streamwise location $x/h_o = 10$.}
    \label{fig:lateralVG_q}
\end{figure}

First, we consider how the lateral position of the vortex impacts control authority. Three positions were tested: (1) 6.75 orifice widths left of the jet, (2) centered, and (3) 6.75 orifice widths right of the jet; which allows us to test the jet effectiveness on both the upward and downward velocity sides of the vortex where the vortex is primarily outside of the jet width. Figure \ref{fig:lateralVG_q} shows the vortex field of the baseline case compared to the actuated case for the three lateral positions measured at $x/h_o=10$. In all three positions, there is a strong impact on the vortex due to the jet (even within a few orifice lengths downstream). However, the case where the vortex is left of the orifice has lowest effect, the core of the original vortex was still intact. In this case, the velocity field of the jet may be directly assisting the upward velocity side of the vortex, having an undesired effect as the wall-normal jets ($\alpha = 90^\circ$) show strong vertical velocities in this region. When the vortex is centered and right of the orifice we see similar levels of vortex destruction, with the vortex on the right being the most impacted. This is likely because on this side of the vortex, the upward velocity from the jet counters the downward velocity of the vortex---opposite the effect seen when the vortex was on the other side of the orifice.  For all cases, we can see smaller secondary vortex structures generated, either counter- or co-rotating with the original vortex. 

\begin{figure}
    \centering
    \includegraphics[width=0.93\linewidth]{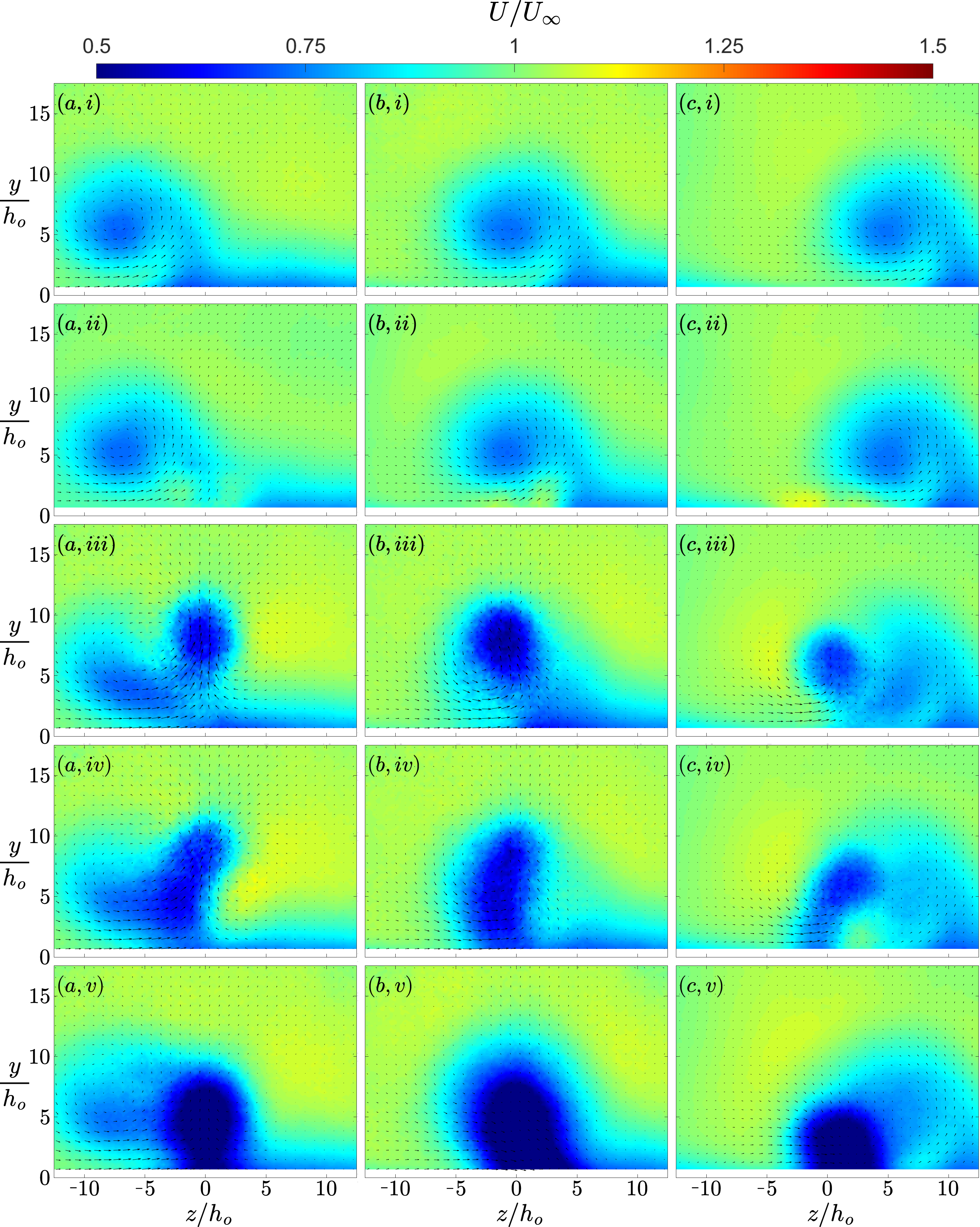}
    \caption{Velocity flow fields showing the impact of lateral vortex position on control effectiveness for a vortex positioned (a) $z/h_o = -6.75$, (b) $z/h_o = 0$, and (c) $z/h_o = 6.75$ relative to the synthetic jet centroid. Baseline flow fields of just the vortex generators are shown in row (i). Synthetic jet orientations were (ii) $\alpha = 45^\circ$ \& $\beta=0^\circ$, (iii) $\alpha = 90^\circ$ \& $\beta=0^\circ$, (iv) $\alpha = 90^\circ$ \& $\beta=30^\circ$, and (v) $\alpha = 135^\circ$ \& $\beta=0^\circ$. All flow fields shown are at streamwise location $x/h_o = 10$.}
    \label{fig:lateralVG_velo}
\end{figure}

\begin{figure}
    \centering
    \includegraphics[width=0.93\linewidth]{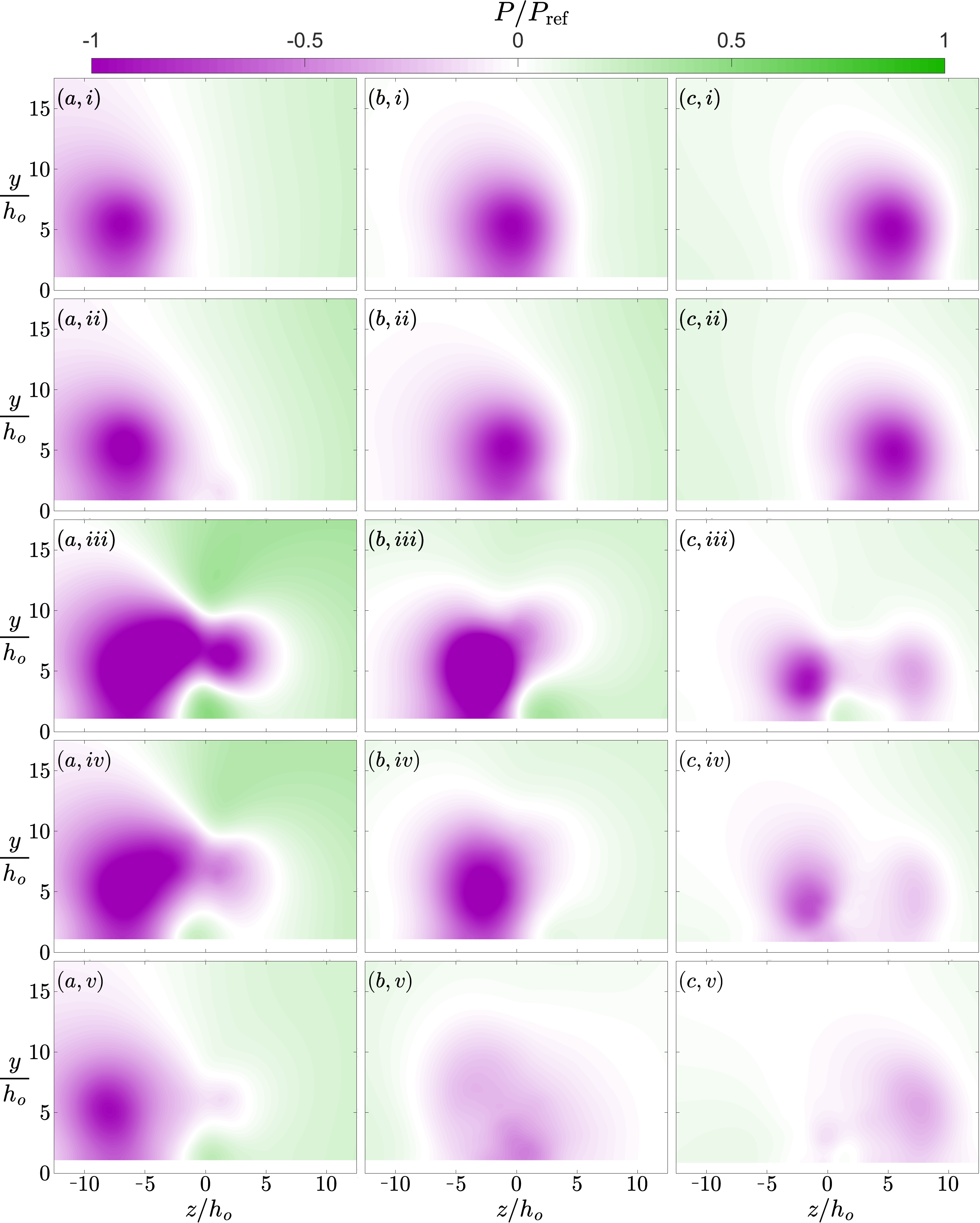}
    \caption{Pressure fields showing the impact of lateral vortex position on control effectiveness for a vortex positioned (a) $z/h_o = -6.75$, (b) $z/h_o = 0$, and (c) $z/h_o = 6.75$ relative to the synthetic jet centroid. Baseline flow fields of just the vortex generators are shown in row (i). Synthetic jet orientations were (ii) $\alpha = 45^\circ$ \& $\beta=0^\circ$, (iii) $\alpha = 90^\circ$ \& $\beta=0^\circ$, (iv) $\alpha = 90^\circ$ \& $\beta=30^\circ$, and (v) $\alpha = 135^\circ$ \& $\beta=0^\circ$. All flow fields shown are at streamwise location $x/h_o = 10$.}
    \label{fig:lateralVG_Pressure}
\end{figure}

The velocity field for the same cases as in figure \ref{fig:lateralVG_q} are shown in figure \ref{fig:lateralVG_velo}. For all three lateral vortex positioning cases, there are strong wake regions of the jet that combine with the wakes of the vortex to varying degrees. When the vortex is to the left or right of the orifice, the two wakes are somewhat distinct, except for row (ii). The case with the centered vortex has the strongest and most prominent combined wake. When the vortex is to the right of the orifice, the overall wake structure is lessened compared to the other actuated cases. This vortex position also corresponded to the best vortex destruction position as seen in figure \ref{fig:lateralVG_q}, suggesting simultaneous vortex destruction and wake reduction is possible. The jet pitched in the crossflow direction $\alpha=45^\circ$ \& $ \beta = 0^\circ$, figure \ref{fig:lateralVG_velo}(a-c, ii), is the only jet orientation to reduce velocity deficit. All other cases have a larger velocity deficit than the baseline. For completeness, we also display the corresponding pressure fields calculated from the velocity in figure \ref{fig:lateralVG_Pressure}. Overall, we see a continuing trend that vortex structure diminishment correlates to pressure recovery, with the right-most vortex position recovering best.

\begin{figure}
    \centering
    \includegraphics[width=\linewidth]{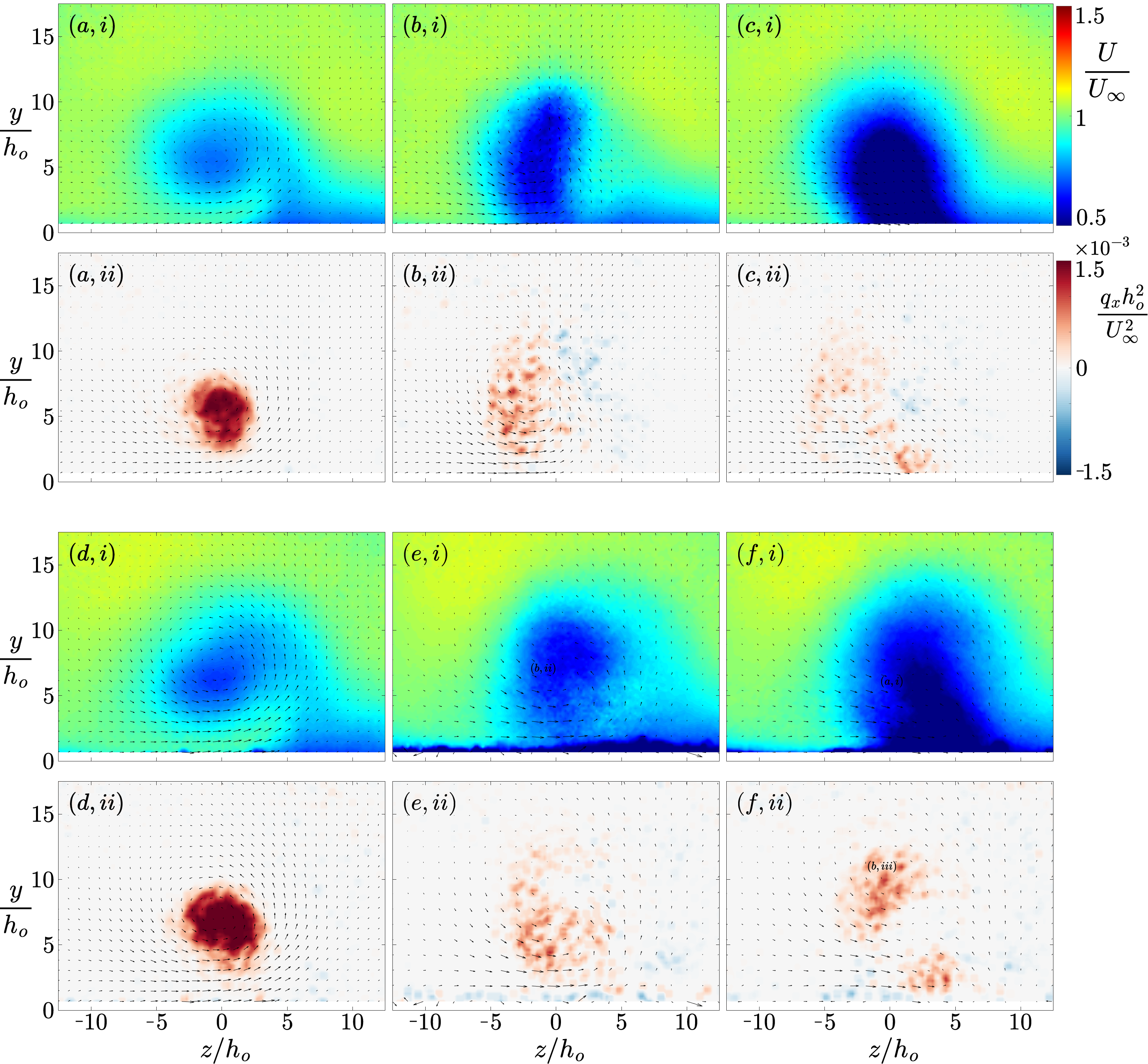}
    \caption{Velocity (i) and vortex (ii) field showing the impact of vortex size on control effectiveness for a vortex with (a-c) $r_{vg} = 3.75h_o$ and (d-f) $r_{vg} = 5 h_o$ comparing the baseline (a/d) and actuated cases with orifice orientation (b/e) $\alpha = 90^\circ$ \& $\beta = 30^\circ$ and blowing ratio $C_b = 1.3$ for the $\alpha = 90^\circ$ jet and $C_b = 1.0$ for pitched (c/f) $\alpha = 135^\circ$ jet, measured at $x/h_o = 10$.}
    \label{fig:vortexSize}
\end{figure}

After considering lateral positioning of the vortex, we can now move to the impact of vortex size on the flow control effectiveness. The baseline cases showing just the vortex generator, along with our two best performing jet orientations are shown in figure \ref{fig:vortexSize}. The vortices were produced by similar vortex generators where only the height was varied, either $h_{VG}/h_o = 3.75$ or 5, and the vortex generator was in the central position. This produces a vortex approximately 33\% larger than the one we have used for the majority of this study. Despite increasing the vortex size, all cases produce a similar diminished vortex coherence compared to the baseline. Qualitatively, the jet pitched into the crossflow ($\alpha = 135^\circ$) was better at breaking down the smaller vortex, while the jet oriented $\alpha = 90^\circ$ \& $\beta = 30^\circ$ was more effective with the larger vortex. Overall, without modifying the size or strength of the synthetic jets, the jets were able to diminish the vortex coherence for not only our representative $h_{VG}/h_o = 3.75$ case but also our larger $h_{VG}/h_o = 5$ vortex generator. We do recognize that this is not a wide parameter space of vortex sizes, and there must be an upper limit on how big a vortex can be before the jet is not effective. However, it is encouraging that we are altering vortices that are around the same height as the local boundary layer, $0.8-1.0 \times \delta_\text{99}$.

\begin{figure}
    \centering
    \includegraphics[width=1\linewidth]{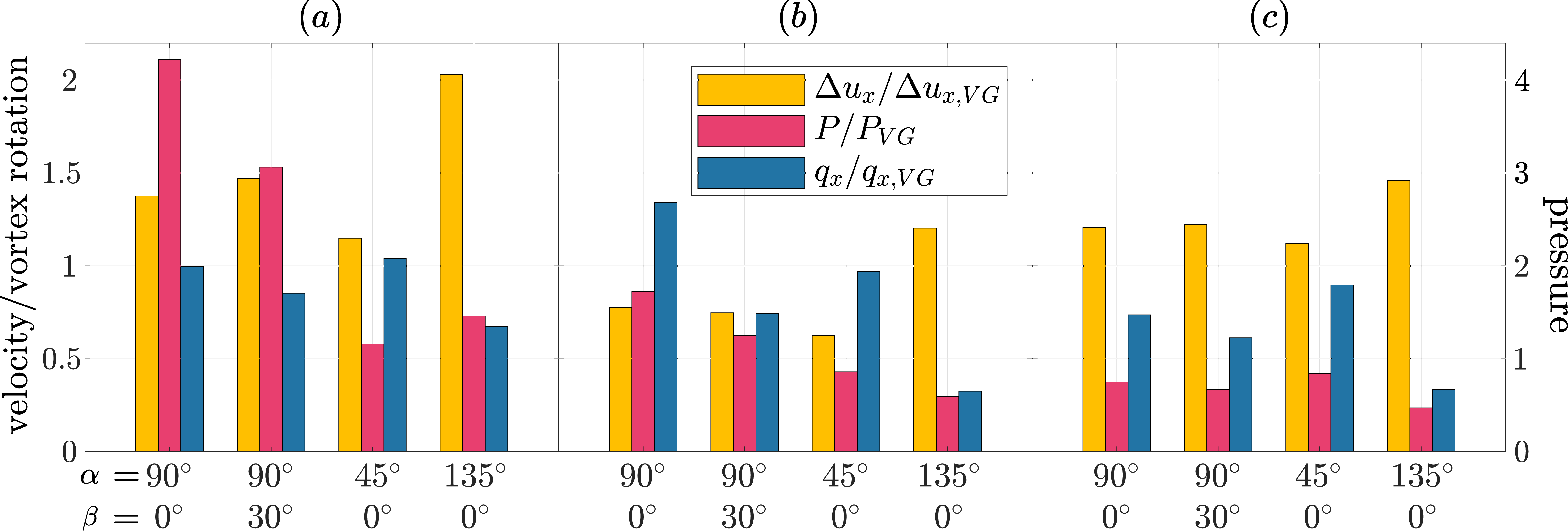}
    \caption{Coefficient bar graphs for all jet orientations at each lateral position (a) $z/h_o = -6.75$, (b) $z/h_o = 0$, and (c) $z/h_o = 6.75$ relative to the synthetic jet centroid. All values are from the streamwise location $x/h_o = 10$ and are normalized by their corresponding vortex generator with height $h_{VG}/h_o = 3.75$. Note, the change in streamwise velocity and vortex structure bars are referenced to the left y-axis, while the pressure bars are plotted with respect to the right y-axis.}
    \label{fig:BarGraphs}
\end{figure}

Finally, we look to quantify the effectiveness of the flow control with respect to the vortex rotation, velocity wake, and pressure recovery.  We area-average these three fields at a fixed downstream location $x/h_o = 10$ for all three lateral vortex positions and all four orientations in figure \ref{fig:BarGraphs}. We normalize each quantity by the respective isolated vortex generator baseline case. To calculate the change in streamwise velocity ($\Delta u_{\cdot}$), we subtract the local flow no jet or vortex generator. All quantities are area-averaged between $z/h_o = \pm 12.5$ and $y/h_o = 2.5$ to 20. 

First, we can look at the vortex structure $q_x/q_{x,VG}$, figure \ref{fig:BarGraphs}. Generally, the rotational structure goes down as the vortex position moves from the left side (where the upward flow side of the vortex can be assisted by the jet) to the right side of the orifice (where the downward side of the vortex is against the jet). This trend would almost certainly not continue as the vortex would eventually leave the jet proximity. With the vortex in the ideal lateral position, all jet orientations reduce the rotation from 20-70\% reduction. For the other two cases, we still see largely see a global reduction in strength, except for $\alpha = 90^\circ$ and $\beta = 0^\circ$ in the middle vortex position. Across all positions, the jet pitched into the crossflow $\alpha = 135^\circ$ and $\beta = 0^\circ$ caused the greatest decrease in vortex structure.

The change in streamwise velocity $\Delta u_x/\Delta u_{x,VG}$ highlights how the velocity deficit region varies with the different jet orientations. Because the vortex generator on its own produces a velocity deficit region, the area average of this region is a negative value. When using this to normalize the jet cases, values greater than 1 indicate a more negative region and values of less than 1 stem from either reducing that deficit region or accelerating the flow. The central vortex position \ref{fig:BarGraphs}(b) is the only position where flow speed increased relative to the baseline vortex generator case. This occurs not just for the jet pitched more along the crossflow $\alpha = 45^\circ$ and $\beta = 0^\circ$ case, but also for the two wall-normal $\alpha = 90^\circ$ jets. As expected, the jet pitched into the crossflow produces a larger deficit region than the other orientations, regardless of vortex position. While the right vortex position $z/h_o = 6.75$ leads to stronger vortex destruction when looking at the vortex structure, the central position $z/h_o = 0$ might be more favorable in some instances due to the added flow momentum here. 

Finally we can look at the effect on the pressure field $P/P_{VG}$. We can directly compare figure \ref{fig:BarGraphs}(c) to the pressure contour plot from before, figure \ref{fig:PressureFieldDifference}. By inspection, we expected each of the four orientations to reduce the overall suction pressure, which is borne out by the statistics. This rightmost vortex position resulted in the greatest reduction in suction pressure, ranging from 15-55\% reduction. For the middle vortex position $z/h_o = 0$, we see a range in effectiveness from almost a doubling in pressure at $\alpha = 90^\circ$ and $\beta = 0^\circ$ to about halving the pressure at $\alpha = 135^\circ$ and $\beta = 0^\circ$. There were no favorable changes to the pressure field when looking at the leftmost vortex generator position, figure \ref{fig:BarGraphs}(a). Interestingly, there were scenarios where the vortex pressure deficit was enhanced by the presence of the jet---reflecting how sensitive this control technique may be to vortex meandering. 

\section{Conclusion}\label{SEC:Conclusion}

Synthetic jets were used to mitigate wall-bounded streamwise vortex structures in a flat plate laminar boundary layer. Rectangular tab-style vortex generators were installed upstream of the synthetic jet orifice in a wind tunnel and measured using stereoscopic particle image velocimetry at multiple streamwise locations. We varied the vortex size and lateral position with respect to the jet with multiple jet orifice orientations, including wall-normal skewed/unskewed as well as pitched into and away from the flow. The vortex was a similar size to the jet orifice length and 0.8-1.0$\times$ larger than the baseline boundary layer height. The jet blowing ratio was fixed at $C_b = 1.0$ or $1.3$ for pitched and wall-normal orifices, respectively. The primary goal of the work was to disrupt the rotational structure of the vortex using the unsteady jet, as this is what leads to the low pressure concentrations that induce often undesired aerodynamic forces in a variety of applications. The jet rotation was captured via a modified $\mathcal{Q}$-criterion and the pressure field was calculated from the velocity field. A secondary consideration was the velocity deficit wake of the vortex.

The majority of the cases tested led to marked decreases in jet rotational coherence, with up to 70\% reduction, leading to pressure recovery within the vortex. The most robust case for vortex destruction was when the jet was angled directly against the flow, $\alpha = 135^\circ$ and $\beta = 0^\circ$, which had the largest blockage and flow disruption. Other cases, like the wall-normal $\alpha = 90^\circ$  skewed $\beta = 30^\circ$ jet, were nearly as effective with much less downstream disturbance. Most jet orientations had little impact on the vortex's velocity wake (which is somewhat expected as the jets produce a wake of their own due to the virtual blockage). However, the orifice pitched along the flow direction, $\alpha = 45^\circ$ and $\beta = 0^\circ$, was capable of moderately accelerating the wake though was less effective at rotational control.

The synthetic jet was capable of disrupting both the small and large vortices tested, though it was sensitive to the lateral position of the incoming vortex. Ideally, the vortex passes directly over the orifice, or favors the orifice side where the vertical velocity of the jet can counter the downward velocity side of the vortex. When the vortex passes toward the other side of the orifice, the jet vertical velocity appears to assist the vortex---a strategy more suitable for vortex winding \cite{remneff2024control}. Ultimately, this work provides evidence that the synthetic jet is a suitable flow control actuator for vortex destruction. Looking forward, we hope to expand these studies to consider a wider range of vortex/jet strengths and sizes as well as moving to lifting surfaces to explore the direct change in global forces due to vortex control.

\section{Acknowledgments}\label{SEC:Acknowledgements}
This work was supported by AFOSR research grant FA9550-22-1-0438. Frank A. Tricouros was the recipient of the DoD SMART Scholarship.

\bibliographystyle{unsrt}   
\bibliography{bibliography}

\end{document}